\documentstyle[prl,aps,preprint,floats]{revtex}
\input psfig.tex

\begin{document}
\draft
\twocolumn[\hsize\textwidth\columnwidth\hsize\csname @twocolumnfalse\endcsname
\title{
Measuring Cosmological Parameters with Cosmic Microwave Background
	Experiments
}
\author{
J. Richard Bond,$^1$ Robert Crittenden,$^2$
Richard L. Davis,$^2$
George Efstathiou,$^3$ Paul J. Steinhardt$^2$}
\address{
$^{(1)}$  Canadian Institute for Theoretical Astrophysics,
University of Toronto,
Toronto, Ontario, Canada
M5S 1A7  \\
$^{(2)}$  Department of Physics, University of Pennsylvania,
Philadelphia, PA  19104 \\
$^{(3)}$  Department of Physics,
 Oxford University, Oxford, England  OX1 3RH }
\maketitle
\begin{abstract}

The cosmic microwave background anisotropy is sensitive to the slope
and amplitude of primordial energy density and gravitational wave
fluctuations, the baryon density, the Hubble constant, the
cosmological constant, the ionization history, {\it etc.}  In
this Letter, we examine the degree to which these factors can be
separately resolved from combined small- and large-angular scale
anisotropy observations.  We isolate directions of degeneracy in this cosmic
parameter space,
 but note that other cosmic observations
can break the degeneracy.
\end{abstract}
]

The observation of large-angular scale ($\sim10^{\circ}$) fluctuations
in the Cosmic Microwave Background(CMB)\cite{dmr,mit} marks the beginning
of a new age of precision measurement in cosmology
\cite{msam,ten,gaier,sask,ovro,max,pwd,python}.  Dramatic
improvements in large- and small-angular scale ($\lesssim 1^{\circ}$)
experiments\cite{msam,ten,gaier,sask,ovro,max,pwd,python} are
anticipated.  In this Letter, we explore the degree to which the CMB
anisotropy observations can determine cosmological parameters
such as the slope of the initial power spectrum, the age of the universe and
the cosmological constant.
We find that CMB anisotropy  measurements alone
cannot fix the parameters individually;
however,
a non-trivial
combination of them can be determined.
More concretely, for models based on
 the generation of gaussian, adiabatic fluctuations by inflation,
we have identified a new variable $\widetilde n_s$,
a function of the basic parameters that can be fixed to great
precision by
CMB anisotropy observations.
Distinct models with nearly the same value of $\widetilde n_s$ cannot
be discriminated by CMB data alone.
In a likelihood analysis, this
leads to
error contours centered around a highly elongated maximum-likelihood
surface
inside which $\widetilde n_s$ is
approximately constant.  However, when
combined with
other, independent  cosmological observations, the determination of
$\widetilde n_s$ is  a powerful tool for
testing models and measuring fundamental parameters.

We parameterize the space by
$$
(C_2^{(S,T,Is,\ldots)},  \;
   n_{s,t,is,\ldots},\; {\rm h},\; \Omega_B,  \; \Omega_{\Lambda},
\Omega_{CDM},
\;  \Omega_{HDM}, \ldots) \ ,
$$
where $H_0 =
100 \, {\rm h} \, {\rm km \, sec^{-1} Mpc^{-1}}$ is the Hubble parameter,
 and $\Omega_{B,\Lambda, CDM,
HDM,\ldots}$ are the energy densities associated with baryons,
cosmological constant ($\Lambda$), cold and hot dark matter, {\it
etc.}, divided by the critical density.
We use the CMB
quadrupole moments $C_2^{(S,T,Is,\ldots)}$ to parameterize the overall
amplitudes of
energy density (scalar metric), gravitational wave (tensor metric),
 isocurvature scalar and other primordial fluctuations predicted by the model.
We parameterize the shape of the initial ({\it e.g.}, post-inflation)
fluctuation spectra in wavenumber $k$
by power law indices $n_{s,t,is,\ldots}$,
defined at time $t_i$ by $k^3
\langle|\widetilde{(\delta \rho/\rho)}(k, t_i)|^2\rangle \propto k^{n_{S}+3}$
and $k^3 \langle|\widetilde{h}_{+, \times}(k, t_i)|^2\rangle \propto
k^{n_{T}}$, where $\delta \rho/\rho$ and $h_{+, \times}$ are the
amplitudes of the energy density and gravitational wave metric
fluctuations (for two polarizations), respectively.

In this Letter, we restrict ourselves to subdomains of this large
space, in particular to parameters consistent with inflation models of
fluctuation generation.  Inflation produces a flat universe, hence
$\Omega_{CDM}+\Omega_{HDM}+\Omega_{B}+\Omega_{\Lambda} \approx 1$. We
also take $\Omega_{HDM} =0$, but note that, for angular scales $\gtrsim
10^\prime$, the anisotropy for mixed dark matter models with
$\Omega_{CDM} + \Omega_{HDM} \approx 1$ is quite similar to the
anisotropy if all of the dark matter is cold. Given $\Omega_B$, we
impose the nucleosynthesis estimate \cite{bbn}, $\Omega_B {\rm h}^2 =
0.0125$, to determine ${\rm h}$, but also satisfy the globular cluster
and other age bounds,\cite{kolb}
 ${\rm h}\lesssim 0.65$ for $\Omega_{\Lambda}=0$
and ${\rm h} \lesssim 0.88$ for $\Omega_{\Lambda} \lesssim 0.6$.
(Gravitational lens statistics \cite{eturner} suggest
$\Omega_{\Lambda} \lesssim 0.6$. A straightforward match to galaxy
clustering data gives $\Omega_\Lambda \le 1- (0.2\pm 0.1){\rm h}^{-1}$
if $n_s \le 1$.\cite{ebw})

Inflation produces adiabatic scalar \cite{G4} and tensor \cite{AWSta}
Gaussian fluctuations.  (For simplicity, we do not consider
isocurvature fluctuations\cite{eb}.)  The COBE quadrupole fixes
$C_2^{(T)}+C_2^{(S)}$, but the tensor-to-scalar quadrupole ratio
$r\equiv C_2^{(T)}/C_2^{(S)}$ is undetermined
\cite{Dav}.  Inflation does not produce strict power-law spectra, in
general, but $n_s$ and $n_t$ can be defined from power-law best-fits
to the theoretical prediction over the scales probed by the CMB.
For generic models of inflation,
including new, chaotic,
and extended models, inflation gives\cite{Dav,Oth,crit}
\begin{equation}
    n_t \approx n_s-1 \; \; {\rm and}
\; \; r \equiv  C_2^{(T)}/C_2^{(S)} \approx 7 (1-n_s) \ .
\end{equation}
Measuring $r$ and $n_s$ to determine whether they respect  Eq.~(1)
is a critical test for inflation.
With this set of assumptions, we  have reduced the parameter-space to
three-dimensions, $(r | n_s,{\rm h}, \Omega_{\Lambda})$ (where
$\Omega_B =0.0125 h^{-2}$ and $\Omega_{CDM} = 1-
\Omega_B-\Omega_{\Lambda}$).  We explicitly display both $r$ and $n_s$
but with a ``$|$" as a reminder that $r$ is determined by Eq.~(1)
given $n_s$; we have also assumed $n_t=n_s-1$.

Our results are based on numerical integration of the general
relativistic Boltzmann, Einstein, and hydrodynamic equations for both
scalar\cite{BE} and tensor metric fluctuations using methods reported
elsewhere\cite{crit}.  Included in the dynamical evolution are all the
relevant components: baryons, photons, dark matter, and massless
neutrinos.  The temperature anisotropy, $\Delta T/T\, (\theta , \phi)=
\sum_{\ell m} a_{\ell m} Y_{\ell m} (\theta,\phi)$, is computed in
terms of scalar and tensor multipole components, $a_{\ell m}^{(S)}$
and $a_{\ell m}^{(T)}$, respectively.  For inflation, each multipole
for the two modes is predicted to be statistically independent and
Gaussian-distributed, fully specified by angular power spectra,
$C_\ell^{(S)}=\left\langle \vert a_{\ell m}^{(S)}\vert^2
\right\rangle$ and
$C_\ell^{(T)}=\left\langle \vert a_{\ell m}^{(T)}\vert^2
\right \rangle$.

\begin{figure}[t1]
\caption{Top: Power spectra as a function of multipole moment $\ell$
for ($r$=$0 | n_s$=$1$), ($r$=$0.7 | n_s$=$0.9$) and ($r$=$1.4 | n_s$=$0.8$)
where
${\rm h}=0.5 \; {\rm and} \; \Omega_{\Lambda}=0$ for all models.  The
spectra in all figures are normalized by the COBE
$\sigma_T^2(10^\circ)\equiv (4 \pi)^{-1}\sum (2 \ell +1) C_l
\, {\rm exp} (- \ell(\ell+1)/158.4)$.
(a Gaussian filter with $10^\circ$
{\it fwhm}), observed by DMR to be $\sim 1.2 \times 10^{-10}$, with
about a 30\% error.  Bottom: $(\Delta T/T)_{rms}$ levels with 1-sigma
cosmic variance error bars for nine experiments assuming full-sky
coverage.  [For $N_D=50$ patches and a unity signal-to-noise ratio,
the variance is $20\%$; see Eq.~(2)].  The gaussian coherence angle
 is indicated below each experiment; see Refs.
1-11 for acronyms. }
\end{figure}

\begin{figure}[t2]
\caption{Power spectra as a function of $\ell$ for
scale-invariant models, with $r=0 | n_s=1$.  The middle curve shows
${\rm h}=0.5$ and $\Omega_{\Lambda} =0$. In the upper
curve, $\Omega_{\Lambda}$ is increased to 0.4 while keeping ${\rm
h}=0.5$.  In the lower  curve, $\Omega_{\Lambda}=0$ but ${\rm
h}$ is increased from 0.5 to 0.65 (hence $\Omega_B$ drops from $0.5$ to
$0.3$).  The spectra are insensitive to changes in ${\rm h}$ for fixed
$\Omega_B$. Increasing $\Omega_{\Lambda}$ or $\Omega_B$ increases the
power at $\ell \sim 200$.  }
\end{figure}

Our results are presented in a series of two-panel figures ({\it e.g.}
see Fig.~1).  The upper plots show the spectrum $C_{\ell}$'s
normalized to COBE, and the lower bar charts show the predicted
$(\Delta T/T)_{rms}$ for idealized experiments spanning $~10^{\circ}$
to $2'$.
The bar chart is constructed by computing $\left\langle(\Delta
T/T)^2\right\rangle =\frac{1}{4 \pi} \sum(2\ell+1)C_\ell W_\ell$, where
$W_\ell$
is a filter function that quantifies experimental
sensitivity.\cite{crit,belm} Errors arise from experimental noise and ``cosmic
variance'', the latter a theoretical uncertainty due to observing the
fluctuation
distribution from only one vantage point.  The errors
bars represent cosmic variance alone assuming full-sky coverage,
exemplifying the limiting resolution achievable with CMB experiments.
For more realistic error bars, consider a detection obtained from
measurements $(\overline{\Delta T/T})_i \pm \sigma_D$ (where
$\sigma_D$ represents detector noise) at $i=1,\ldots
,N_D$ experimental patches sufficiently isolated from each other
to be largely uncorrelated.
 For large $N_D$, the likelihood function falls by
$e^{-\nu^2 /2}$ from a maximum at $(\Delta T/T)_{max}$ when
\begin{equation}  \label{var}
\left(\frac{\Delta T}{T}\right)^2 = \left(\frac{\Delta T}{T}\right)_{max}^2
 \pm \sqrt{\frac{2}{N_D}}\, \nu \,  [\left(\frac{\Delta T}{T}\right)_{max}^2 +
    \sigma_D^2] \, .
\end{equation}
An experimental noise $\sigma_D$ below $10^{-5}$ is standard now, and
a few times $10^{-6}$ is soon achievable, hence if systematic errors
and unwanted signals can be eliminated, the 1-sigma ($\nu =1$) relative
uncertainty in $\Delta T/T$ will be from cosmic-variance alone,
$1/\sqrt{2N_D}$, falling below $10\%$ for $N_D > 50$. The optimal variance
limits shown in the figures roughly correspond to filling the sky with
patches separated by $2\theta_{fwhm}$.

Figure 1 shows a sequence of spectra with varying $r | n_s$.  The
characteristic feature is increasingly suppressed small-angular signal
as $r$ increases and $n_s$ decreases.\cite{Dav,crit} Although cosmic
variance is significant for large-angle experiments,\cite{silk} it can
shrink to insignificant levels at smaller scales if large maps are
made.  It appears that $r | n_s$ would be experimentally resolvable if
$\Lambda$, ${\rm h}$ and ionization history were known.

\begin{figure}[t3]
\caption{Examples of different cosmologies with nearly
identical spectra of multipole moments and $(\Delta T/T)_{rms}$.  The
solid curve is $(r=0 | n_s=1,{\rm h}=0.5, \Omega_{\Lambda}=0)$.  The
other two curves explore degeneracies in the $(r=0 | n_s=1,{\rm
h},\Omega_{\Lambda})$ and $(r | n_s,{\rm h}=0.5,\Omega_{\Lambda})$
planes.  In the dashed curve, increasing $\Omega_{\Lambda}$ is almost
exactly compensated by increasing ${\rm h}$.  In the dot-dashed curve,
the effect of changing to $r=0.42 | n_s=0.94$ is nearly compensated by
increasing $\Omega_{\Lambda}$ to 0.6.  }
\end{figure}

Figure 2 shows the effects of varying $\Omega_{\Lambda}$ or $H_0$
compared to our baseline (solid
line) spectrum $(r=0 | n_s=1,{\rm h}=0.5,\Omega_{\Lambda}=0)$.
Increasing $\Omega_{\Lambda}$  enhances
  small-angular scale
anisotropy  by
reducing  the red shift $z_{\rm eq}$ at which
radiation-matter equality occurs; increasing ${\rm h}$ increases
 $z_{\rm eq}$ and so has the opposite effect.  Increasing
$\Omega_{\Lambda}$ also
 changes slightly  the spectral slope for $\ell
\lesssim 10$ due to $\Lambda$-suppression of the growth of scalar
fluctuations\cite{kofman86}.
  The bar chart shows that either $r | n_s$,
$\Omega_{\Lambda}$, or ${\rm h}$ can  be resolved  if the
other two parameters are known.

\begin{figure}[t4]
\caption{
Power spectra for models with standard recombination (SR), no
recombination (NR), and `late' reionization (LR) at $z=50$.  In all
models, ${\rm h}=0.5$ and $\Omega_{\Lambda}=0$.  NR or
reionization at $z \ge 150$ results in substantial suppression at
$\ell \ge 100$.  Models with reionization at $20 \le z \le 150$
give moderate suppression that can mimic decreasing $n_s$ or
increasing ${\rm h}$; {\it e.g.}, compare the $n_s=0.95$ spectrum with SR
(thin, dot-dashed) to the $n_s=1$ spectrum with reionization at $z=50$
(thick, dot-dashed). }
\end{figure}

A degree of ``cosmic confusion" arises, though, if $r | n_s$,
$\Omega_{\Lambda}$ and ${\rm h}$ vary simultaneously.  Figure 3 shows
our baseline spectrum and spectra for models lying in a
two-dimensional surface of $(r|n_s,{\rm h},\Omega_{\Lambda})$ which
produce nearly identical spectra.  In one case, $r | n_s$ is fixed,
and increasing $\Omega_{\Lambda}$ is nearly compensated by increasing
${\rm h}$.  In the second case, ${\rm h}$ is fixed, but increasing
$\Omega_{\Lambda}$ is nearly compensated by decreasing $n_s$ (with $r$
given by Eq.~(1)).\cite{kofman92}

Further cosmic confusion arises if we also consider ionization
history.\cite{note1} We expand the parameter-space to include $z_R$,
the red shift at which we suppose sudden, total reionization of the
intergalactic medium.  Fig.~4 compares spectra with standard
recombination (SR), no recombination (NR) and late reionization (LR)
at $z_R=50$, where ${\rm h}=0.5$ and $\Omega_{\Lambda}=0$.  NR
represents the behavior if reionization occurs early ($z_R>>200$).
The spectrum is substantially suppressed for $\ell \gtrsim 200$
compared to any SR models.  Experiments at $\lesssim 0.5^{\circ}$
scale can clearly identify NR or early reionization ($z_R \gtrsim 150$
gives qualitatively similar results to NR).
Reionization for $20 \lesssim z_R \lesssim 150$ results in modest
suppression at $\ell \approx 200$, which can be confused with a
decrease in $n_s$ (see figure).

The results can be epitomized by some simple rules-of-thumb:
Over the $30'-2^{\circ}$ range,   $(\Delta T/T)_{\rm rms}^2$ is
roughly proportional to the maximum of $\ell(\ell+1)C_{\ell}$
(the first Doppler peak).  Since
the maximum (corresponding to $\sim .5^{\circ}$ scales) is normalized
to COBE DMR (at $\sim 10^{\circ}$), its value is exponentially
sensitive to $n_s$.  Since scalar fluctuations account for the maximum,
the maximum decreases as $r$ increases.   The maximum is also
sensitive to the red shift at  matter-radiation equality (or,
equivalently, $(1-\Omega_{\Lambda})h^2$), and to the optical
depth at last scattering for late-reionization models, $\sim z_R^{3/2}$.
These observations are the basis of an empirical formula
(accurate to $\lesssim 15$\%)
\begin{equation}
\frac{\ell(\ell+1)C_{\ell}}{2 \pi \sigma_T^2(10^{\circ})}\Big\vert_{max}
\approx A \; e^{B \; \tilde{n}_s}
\end{equation}
where $A=0.1$, $B=3.56$, and
 \begin{equation}
\begin{array}{rcl}
\tilde{n}_s &  \approx &  n_s -0.28 \, {\rm log}(1+ 0.8 r) \\ & &
-0.52 [(1-\Omega_{\Lambda})h^2]^{\frac{1}{2}}\,
    - 0.00036 \, z_R^{3/2}+.26 \ ,
\end{array}
\end{equation}
where $r$ and $n_s$ are related by Eq.~(1) for generic inflation models,
 and $z_R \lesssim 150$ is needed to have a local maximum.
($\tilde{n}_s$ has been defined such that $\tilde{n}_s=n_s$ for $r=0,\;
h=0.5, \; \Omega_{\Lambda}=0,$ and $z_R=0$.)

Our central result is that
CMB anisotropy experiments can determine $\tilde{n}_s$, but
variations of parameters along the surface of constant $\tilde{n}_s$
produce indistinguishable CMB anisotropy.
Given present uncertainties in
${\rm h}$, $\Omega_{\Lambda}$ and $z_R$, it will be possible to
determine the true spectral index $n_s$ (or $r$) to within
$~10\%$ accuracy
using the CMB anisotropy alone.  Quantitative improvement can
be gained by invoking constraints from large-scale structure, {\it
e.g.}, galaxy velocity and cluster distributions, although the results
are model-dependent.  Ultimately, tighter limits on
$\Omega_{\Lambda}$, ${\rm h}$, ionization history, and the dark matter
density are needed before the CMB anisotropy can develop into a high
precision test of inflation (Eq.~(1)) and primordial gravitational
waves.

This research was supported by the DOE at Penn (DOE-EY-76-C-02-3071), NSERC
at Toronto, the SERC at Oxford and the Canadian Institute for Advanced
Research.


\begin{references}
\bibitem{note1}
Inflation-based models are likely to have $z_R$ low enough to
have little effect,\cite{BE} but, by imposing it, we can illustrate
the qualitative behavior for models which should have $z_R$ high ({\it
e.g.}, cosmic strings and textures).
\end{references}
\end{document}